\documentclass[10pt,letterpaper]{article}
\usepackage{opex3}

\begin{document}




\title{Twin-Photon Confocal Microscopy}

\author{D.S. Simon$^{1,2,\ast}$ and A.V. Sergienko$^{1,2,3}$}

\address{$^1$ Dept. of Electrical and Computer Engineering, Boston
University, 8 Saint Mary's St., Boston, MA 02215.

$^2$ Photonics Center, Boston
University, 8 Saint Mary's St., Boston, MA 02215.

$^3$ Dept. of Physics, Boston University, 590 Commonwealth Ave.,
Boston, MA 02215.}

\email{$^\ast$ simond@bu.edu} 



\begin{abstract}
A recently introduced two-channel confocal microscope with correlated detection promises up to 50\% improvement
in transverse spatial resolution [Simon, Sergienko, Optics Express {\bf 18}, 9765 (2010)] via the use of photon correlations. Here we achieve similar results in a different manner, introducing a triple-confocal correlated microscope which exploits the correlations present in optical parametric amplifiers. It is based on tight focusing
of pump radiation onto a thin sample positioned in front of a nonlinear crystal, followed by coincidence detection
of signal and idler photons, each focused onto a pinhole. This approach offers further resolution enhancement in confocal microscopy.
\end{abstract}

\ocis{(180.1790) Confocal microscopy; (180.5810) Scanning
microscopy; (190.4970) Parametric oscillators and amplifiers.}

\section{Introduction}

\subsection{Confocal microscopy.}

Confocal microscopy has two main advantages over widefield microscopy. The first is the improvement in contrast
introduced by the pinholes, which block any light that has not passed through a small in-focus region of interest.
The second is an improvement in resolution due to a double passage through the objective lens and the requirement that
the in-focus region for both passages overlap. If the objective is a lens of focal length $f$ and circular aperture of
radius $a$ described by pupil function $p({\mathbf x})$, then the point spread function (PSF) for a confocal microscope is
(up to normalization)
\begin{equation}PSF_{con}({\mathbf x})=\tilde p^4\left({{k{\mathbf x}}\over f}\right) =\left| {{J_1(kax/f)}\over {(kax/f)}}\right|^4
\end{equation} where the tilde represents Fourier transform and $J_1$ is a Bessel function of first order.
This is to be compared to the widefield PSF for the same lens, \begin{equation}PSF_{wf}({\mathbf x})=\tilde p^2\left({{k{\mathbf x}}\over f}\right) .
\end{equation} For the confocal microscope, each passage through the lens gives a factor of $\tilde p^2$; since the light passes through the lens twice, this becomes squared to give the more sharply peaked function $\tilde p^4$. The result is a PSF for the confocal microscope
which is roughly 28\% narrower than that of the corresponding widefield microscope.

\subsection{Multiple photons in confocal microscopy.}

Several types of device have previously made use of multiple photons to improve resolution within the context of confocal microscopy. For example, {\it two photon confocal microscopy} \cite{denk1,denk2} requires two photons to {\it simultaneously excite} the same fluorescent molecule, thus limiting the visible region at each point of the scan to the small volume for which the intensity
is large enough for two excitations to occur simultaneously with reasonable probability.  Recently, {\it correlation confocal microscopy} \cite{simon} has been proposed, in which the photon pairs are {\it simultaneously detected}, with a detection scheme designed such that only pairs that struck the sample within a small distance of each other have a high probability of being detected.

\begin{figure}
\centering
\includegraphics[totalheight=2in]{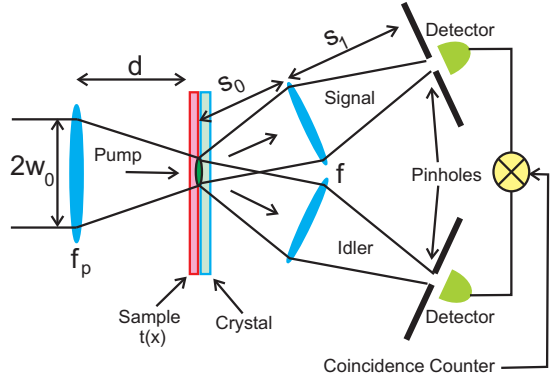}
\caption{\textit{(Color online) Twin photon microscope}}\label{fig1}
\end{figure}

The goal here is to go one step beyond confocal microscopy by requiring the overlap of {\it three} in-focus regions, two detection and one illumination region (see fig. 1). Only points in the overlap of all three regions will be visible. If the third region is comparable to the size
of the objective's Airy disk, the overall area that can be resolved will become smaller due to the combined drop-off of the detection and illumination probabilities. This will be achieved by means of
spontaneous parametric downconversion (SPDC). The two outgoing photons (signal and idler)
each pass through an objective lens of focal length $f$, and then through
pinholes to photon-counting detectors. These signal and idler branches are analogous to the two halves of a standard confocal microscope. The
detectors will be connected in coincidence, so that detection events will be recorded only when {\it both} of the photons survive the corresponding
pinhole. This alone would give us the characteristic $\tilde p^4$ behavior of a confocal microscope. To go further, we must look at the
pump beam. Typically, the pump is an approximate plane wave over an area of size much larger than the Airy disk of a microscope lens.
However, it is possible to focus
the pump beam to a much smaller region by means of a lens of focal length $f_p$. The effect of weak focusing of the pump beam in SPDC was
studied in ref. \cite{pittman96}, where it was shown that interesting geometrical optics effects could be obtained, with the
pump beam acting like a spherical mirror, and the signal and idler acting like incident and reflected beams from the mirror.
Here, we work within the framework of \cite{pittman96}, but study the effect of {\it strongly} focusing the pump to a point-like region, and show that the effective overlap region of the signal, idler, and pump can be made small enough to noticeably enhance the resolution over that of the standard confocal microscope. In the process, we will
find that the relationship of the pump beam to the signal and idler causes a second effect that introduces an additional significant spatial resolution enhancement in the lateral direction. The result is that visible-light images can be produced with resolution that normally would be possible only in the ultraviolet.

The device we propose here shares
a common philosophy with two-photon and correlation confocal microscopy, using multiple photons for each detection; but it uses a new mechanism to arrange this. Here, we make intrinsic
use of three photons (signal, idler, and pump), although we only detect two. The constraints involved in parametric downconversion
(energy and momentum conservation, etc.) give us full
knowledge of the properties of all three photons from the measurement of any two of them. Note that
all three of these microscopies use spatial correlations between photons, but that the correlations are imposed at different points: (i) In two-photon microscopy the correlation between the photons is imposed at the {\it sample},
by means of the molecule-photon interaction. (ii) In correlation confocal microscopy the photon correlation is imposed at the {\it detection} stage by a form of postselection. (iii) In the twin-photon confocal microscope proposed here the
correlations between detected photons are imposed at their {\it source}, the downconversion process. We exploit the fact that downconversion is a highly localized process: if the pump is constrained to be inside a very small region near the axis, then its twin daughters must have arisen within the same region.

In contrast to standard two-photon microscopy, the twin-photon microscope can operate with a continuous wave laser source, rather than requiring pulsed lasers. In addition, parametric downconversion provides an entangled-photon source which guarantees that the two photons in a pair are always produced and detected simultaneously; in contrast, standard two-photon microscopy relies on the chance occurrence of two photons arriving simultaneously at the molecule. As a result of these facts and of the improved signal-to-noise ratio induced by the coincidence detection, the twin-photon microscope can operate at lower illumination intensities than the standard two-photon microscope, leading to lower power requirements and less damage to the sample.

\subsection{Resolution, the Abbe limit, and blurring by material.}

As the resolution in an optical system improves, eventually it runs up against the Abbe diffraction limit. Beyond this point, imaging the shape of smaller features in extended objects with an ordinary microscope is not possible since the high spatial-frequency components needed to reconstruct the shape will not propagate into the far field or diffract into the objective lens. Thus, it could be argued that beyond a certain point it becomes meaningless to seek further improvements in the resolution of a confocal microscope. However, the derivation of the Abbe limit makes use of two assumptions: (i) uniform illumination and (ii) linear response in the detection system. When these two assumptions are removed, it is possible to expand the range of spatial frequencies the system allows to pass, leading to superresolution \cite{lukosz}. This fact has been exploited in a number of different types of fluorescence-based far-field microscopes in recent years (see \cite{heintzmann} for an overview). Here, both of the listed assumptions are violated: we have nonuniform illumination (pump beam focused to a spot) and nonlinear response in the detection system (the coincidence detection leads to quadratic response). Thus the usual Abbe limit does not apply, allowing an expansion of the spatial frequency pass-band similar to those obtained with the methods described in \cite{heintzmann}, but achieved in our case using a non-fluorescence-based far-field microscope.

Moreover, light will scatter even from very small particles, including those of extreme sub-wavelength size. Even though the high spatial-frequency components needed to reconstruct the object's shape may be missing from the image, this scattered light may still be used to identify the presence of such particles and to localize their positions. The use of optical microscopes (especially polarization microscopes) for this is an active field of research: a number of papers have appeared, for example, which study the use of polarizing confocal and widefield microscopes to
view subwavelength dielectric particles \cite{torok1,wilson,torok2,torok3,sheppard}. So, it is worthwhile to continue investigations of resolution improvements in far-field microscopes well beyond the Abbe limit.

Note further that in the twin-photon microscope proposed here the combination of pinholes, focused beam, and coincidence detection provide a very high degree of weeding out of stray light from regions outside the area of interest. In particular, multiply-scattered light will not contribute to the signal. Thus, the effect of the surrounding material will be greatly reduced, leading to less blurring of the image. Furthermore, the coincidence method reduces the effect of dark current noise in the detectors, as well as other sources of noise, leading to improved signal-noise-ratio. The chief trade-off is that the reliance on parametric down-conversion and coincidence detection will reduce the counting rate, leading to longer collection times than is required to form an image by standard confocal microscopy.

In the following sections, we describe the apparatus and analyze its behavior. The goal of the current paper is to describe the main ideas as simply as possible. In order to keep the analysis simple we make
use of several approximations. Specifically, we use scalar diffraction theory in the paraxial approximation, and we assume infinitesimal pinholes. However, these approximations can be inappropriate when used in conjunction with the high NA lenses often found in confocal microscopes.
So, in order to verify that the main points of the paper are not much altered by the approximations used,
the authors have repeated the same analysis with nonparaxial vector diffraction theory and pinholes
of finite size taken into account. Using this more exact analysis it has been shown that the resolution enhancement found in the simplified version presented in this paper does indeed survive when the approximations are removed. Including the full vector treatment here would make the current paper too cumbersome and would obscure the relatively simple physical ideas, which are much more apparent in the context of the simplified scalar-diffraction treatment. Therefore, we discuss only the simplified model in the current paper, while the more exact treatment
of the twin-photon microscope using vector fields and finite pinholes will be prepared for separate publication.

\section{The Coincidence Rate and Point Spread Function}

\subsection{Derivation of Coincidence Rate.}

\begin{figure}
\centering
\includegraphics[totalheight=2in]{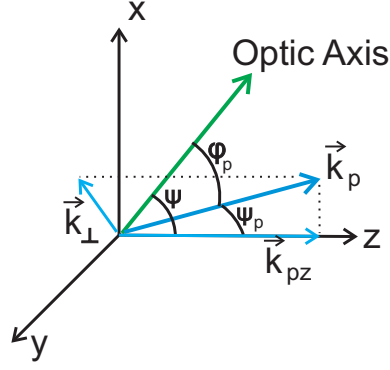}
\caption{\textit{(Color online) Definitions of pump-related angles. Similarly,
the outgoing beams are at angles $\theta_e, \; \theta_o$ from the optic axis.}}\label{opticaxis}
\end{figure}

Consider the setup of fig. \ref{fig1}. The pump beam of waist radius $w_0$ and frequency $\omega_p={{2\pi c}\over \lambda_p}$ is focused by a
lens of focal length $f_p$ to a small region at the face of a $\chi^{(2)}$ nonlinear crystal.
At this face is placed a thin sample of transmittance $t({\mathbf y})$, where ${\mathbf y}$ is the position in
the plane transverse to the propagation direction ($z$). Spontaneous parametric downconversion occurs inside the
crystal, producing two beams, the signal (ordinary ray) and idler (extraordinary ray), of respective frequencies $\omega_o$ and $\omega_e$.
The angles involved in the phase-matching inside the crystal are defined in fig. \ref{opticaxis}. The $z$-axis
is the propagation axis, at angle $\psi$ from the optic axis. 

Without the sample in place, the pump field inside the crystal would be given by \cite{pittman96} \begin{equation}
E_p({\mathbf r_\perp},z,t)=\int d^2k_\perp e^{-i\omega_pt} e^{-i(k_pz +{\mathbf k_\perp}\cdot {\mathbf r_\perp})}
\tilde E_p({\mathbf k_\perp}).\end{equation}
Here, ${k_p}$ and ${\mathbf k_\perp}$ are the pump momentum in the longitudinal and transverse
directions, and ${\mathbf r_\perp}$ is the position within the pump beam in the transverse direction. $\tilde E_p$
is the Fourier transform of the pump field inside the crystal, which is given by
$\tilde E_p({\mathbf k_\perp})=e^{-ik_\perp^2 \sigma_p^2/2} \cite{pittman96},$ with
\begin{equation} \sigma_p^2\approx {c\over {\omega_p}}\left[ d-f_p-i\left( {{\lambda_p}\over {\pi w_0^2}}\right)f_p^2
\right] .\end{equation}  Under normal experimental conditions
the pump field inside the crystal reduces approximately to \cite{pittman96}:
\begin{equation} E_p({\mathbf r_\perp},z,t)= E_p^{\prime \prime } e^{-i(\omega_p t -K_pz)}e^{ir_\perp^2/2\sigma_p^2},
\end{equation} where all overall constants have been swept into $E_p^{\prime \prime }$. Adding the sample multiplies
this at each point by $t({\mathbf r_\perp})$.

Each outgoing beam passes through half of a confocal microscope arrangement: it propagates distance $s_0$, passes
through the objective lens, then propagates another distance $s_1$ before reaching a pinhole. Distances $s_0$ and $s_1$ satisfy
the imaging condition ${1\over {s_0}} +{1\over {s_1}} ={1\over f}.$
If the positive frequency parts of the
outgoing fields of transverse momenta $k^\prime_j$ ($j=e,o$) are $
\hat E_j^{(+)}({\mathbf r_{j\perp}},t) =E_j ({\mathbf r_{j\perp}},t)\hat a_{k^\prime_j},$
then the contribution to the field detected
by a detector located at distance $s_0+s_1$ away
due to field $j$ at radial distance ${\mathbf r_{j\perp}}$ at the crystal
is given by
\begin{equation}\hat E_j^{(+),D}({\mathbf r_{j\perp}},t_j) =\int d^3k_j^\prime h_{k_j^\prime}
({\mathbf r_{j\perp}},t_j)\hat E_j^{(+)}({\mathbf r_{j\perp}},t_j) ,\end{equation} where $h_{k_j^\prime}$
is the amplitude for a ray launched from ${\mathbf r_\perp}$ with momentum $k_j^\prime$ to propagate
through the lenses, survive the pinhole, and reach the detector. For the confocal system shown,
this propagation factor is given up to overall normalization by \begin{equation}h_{k_j^\prime} ({\mathbf
\xi}) = e^{-i(k/s_0)\xi^2} e^{-i{\mathbf {k_\perp}}\cdot {\mathbf \xi}}p\left({\mathbf \xi}+{{s_0}\over k}{\mathbf k_\perp}\right)
.\end{equation}
The contributions to the fields at the detectors due to the downconversion fields at ${\mathbf r_{j\perp}}$
are \begin{equation} E_j^{(+),D}({\mathbf r_{j\perp}},t_j) =\int d^3k_j^\prime
e^{-i(k_j^\prime/s_0)r_{j\perp}^2} e^{-i{\mathbf {k_{j\perp}^\prime}}\cdot
{\mathbf r_{j\perp}}} e^{-i\omega_jT_j}p\left({\mathbf r_{j\perp}}+{{s_0}\over k_j^\prime}{\mathbf k_{j\perp}^\prime}\right)
E_j\hat a_{k_j^\prime},\end{equation} where $T_l=t_j-{{s_0+s_1}\over c}.$ The coincidence amplitude
is  \begin{equation}A(T_1,T_2)=\int d^2r_\perp \langle 0|
E_e^{(+),D}({\mathbf r_{\perp}})E_o^{(+),D}({\mathbf r_{\perp}})|\psi ({\mathbf r_{\perp}},{\mathbf y})\rangle .\label{ampdef}
\end{equation}

The interaction Hamiltonian density for SPDC is \begin{eqnarray}{\cal H}({\mathbf r_\perp },{\mathbf y})&=&
\epsilon_0\chi E_p^{(+)}({\mathbf r_\perp},{\mathbf y})E_e^{(-)}({\mathbf r_\perp})E_o^{(-)}({\mathbf r_\perp})+h.c.\nonumber \\
&=&A_1\int d^3k_ed^3k_o \hat a^\dagger_{k_e}a^\dagger_{k_o}e^{i(\omega_e+\omega_o-\omega_p)t}e^{i(k_p-k_{ez}-k_{oz})z}\nonumber \\
& & \qquad \times \; e^{-i({\mathbf k_{e\perp}}+{\mathbf k_{o\perp}})\cdot {\mathbf r_\perp}}e^{i|{\mathbf r_\perp}|^2/2\sigma_p^2}t({\mathbf r_\perp}
+{\mathbf y}) +h.c., \end{eqnarray} where $A_1$ contains all constants and ${\mathbf y}$ is the displacement of the microscope or sample during scanning.

The two-photon part of the outgoing state vector resulting from downconversion is given by
\begin{eqnarray}|\psi ({\mathbf r_\perp},{\mathbf y})\rangle &=& -{i\over \hbar}\int dt\; dz{\cal H}({\mathbf r_\perp})|0\rangle\\
&=&A_2\int d^3k_ed^3k_o dt\; dz\hat a^\dagger_{k_e}a^\dagger_{k_o}e^{i(\omega_e+\omega_o-\omega_p)t}\label{state}\\
& & \times \; e^{i(k_p-k_{ez}-k_{oz})z}e^{-i({\mathbf k_{e\perp}}+{\mathbf k_{o\perp}})\cdot {\mathbf r_\perp}}e^{i|{\mathbf r_\perp}|^2/2\sigma_p^2}t({\mathbf r_\perp}
+{\mathbf y})|0\rangle .\nonumber\end{eqnarray}
The time integral gives an energy-conserving delta function $\delta (\omega_e+\omega_o
-\omega_p)$, so inserting eq. (\ref{state}) into eq. (\ref{ampdef}): \begin{eqnarray}A(T_1,T_2, {\mathbf y})&=&
A_3 \int d^3k_ed^3k_o d^2r_\perp \int_0^L dz \delta (\omega_e+\omega_o -\omega_p)\;e^{i(k_p-k_{ez}-k_{oz})z}
\nonumber \\ & & \times \; t({\mathbf r_\perp}
+{\mathbf y})p({\mathbf r_\perp}+{{s_0}\over {k_e}}{\mathbf k_{e\perp}})p({\mathbf r_\perp}+{{s_0}\over {k_o}}{\mathbf k_{o\perp}})
e^{i|{\mathbf r_\perp}|^2/2\sigma_p^2}e^{-i(k_o+k_e)r_\perp^2}\nonumber \\ & & \times \; e^{-2i({\mathbf k_{e\perp}}+{\mathbf k_{o\perp}})\cdot{\mathbf r_\perp}} e^{i(\omega_eT_1+\omega_oT_2)}.\end{eqnarray}

We now apply the thin-crystal approximation \cite{pittman96}, in which we may write:
\begin{eqnarray} k_{oz}&=& K_o +{{\nu^\prime }\over {u_o}} -{{|k_{o\perp}|^2}\over {2K_o}} \\
k_{ez}&=& K_e +{\nu\over {u_e}} -N_e|k_{e\perp}|\cos\theta_e +{{|k_{e\perp}|^2}\over {2K_e}}(N_e\cot \psi -1) ,
\end{eqnarray} where we have set $ \omega_o=\Omega_0+\nu^\prime ,$ $\omega_e=\Omega_e+\nu ,$
with $\Omega_o+\Omega_e=\omega_p$. We have also defined \begin{eqnarray} K_o&=& {{\Omega_o}\over c} n_o(\Omega_o), \qquad\quad u_o^{-1} = {d\over {d\Omega_o}}\left[ {{\Omega_o}\over c}n_o(\Omega_o )\right] \\
K_e&=& {{\Omega_e}\over c} n_e(\Omega_e,\psi ),\qquad
u_e^{-1} = {d\over {d\Omega_e}}\left[ {{\Omega_e}\over c}n_e(\Omega_e,\psi )\right] \\
N_e&=& {1\over {n_e(\omega_e,\psi ) }}{d\over {d\psi}} n_e(\omega_e,\psi) .\end{eqnarray}
Furthermore, we may write $\delta (\omega_e+\omega_o -\omega_p)=\delta (\nu +\nu^\prime )$ and decompose the momentum
integration measures according to $d^3k_e
= {1\over c}d^2k_{e\perp}d\nu ,\;
d^3k_o
= {1\over c}d^2k_{o\perp}d\nu $. Making use of all of this, the coincidence
amplitude becomes
\begin{eqnarray}A(T_1,T_2)&=&
A_4 \int d^2k_{e\perp}d^2k_{o\perp} d^2r_\perp d\nu \int_0^L dz e^{-i\nu (T_{12}-Dz)}
e^{-2i({\mathbf k_{e\perp}}+{\mathbf k_{o\perp}})\cdot {\mathbf r_\perp}}\\
& & \times \; e^{-i(\Omega_o+\Omega_e-cs_0/2\sigma_p^2)(r_\perp^2/cs_0)}t({\mathbf r_\perp}
+{\mathbf y})p\left({\mathbf r_\perp}+{{s_0{\mathbf k_{e\perp}}}\over {k_e}}\right) p\left({\mathbf r_\perp}+{{s_0{\mathbf k_{o\perp}}}\over {k_o}}\right) ,\nonumber
\end{eqnarray} where we have once again lumped all overall constants into $A_4$. We have also defined the time difference $T_{12}=T_1-T_2$ and
the inverse group velocity difference $D={1\over {u_o}}-{1\over {u_e}}$. Carrying out the $\nu$ and $z$ integrations, we have
\begin{equation}\int_{-\infty}^\infty d\nu \int_0^L dz \; e^{i\nu (T_{12}-Dz)} =\int_0^L dz\delta (T_{12}-Dz) =\Pi (T_{12}),\end{equation}
where $\Pi (T_{12})$ is the unit step function which is only nonzero for $0<T_{12}<DL$; the presence of  $\Pi (T_{12})$ simply expressed the
fact that both photons must be created simultaneously and within the crystal.
Using $\omega_p=\Omega_o+\Omega_e$, the amplitude now reduces to \begin{eqnarray} A(T_1,T_2,{\mathbf y})&=& A_4 \Pi (T_{12})\int d^2k_{e\perp}d^2k_{o\perp} d^2r_\perp e^{-2i({\mathbf k_{e\perp}}+{\mathbf k_{o\perp}})\cdot {\mathbf r_\perp}}  \\ & & \times \; e^{-ir_\perp^2\left( {{\omega_p}\over {cs_0}}-{1/{2\sigma_p^2}}\right)} t({\mathbf r_\perp}
+{\mathbf y})p({\mathbf r_\perp}+{{s_0}\over {k_e}}{\mathbf k_{e\perp}})p({\mathbf r_\perp}+{{s_0}\over {k_o}}{\mathbf k_{o\perp}})\nonumber  .\end{eqnarray}


We now choose to take $d\approx f_p$, so that \begin{equation}\sigma_p^2\approx -i\left( {{c\lambda_pf^2_p}\over {\pi \omega_pw_0^2}}\right) \equiv -ir_0^2
.\label{r0}\end{equation} $\sigma_p^2$
is now pure imaginary, so that the term $e^{ir_\perp^2/2\sigma_p^2}=e^{-r_\perp^2/2r_0^2}$ in the amplitude decays exponentially with $r_\perp$. Performing a shift of integration variable ${\mathbf k_{j\perp}} \to {\mathbf k_{j\perp}}-{{\Omega_j}\over {s_0c}}r_\perp^2, $ we find the final form of the amplitude:
\begin{equation}A(T_1,T_2,y)= A_5\; \Pi(T_{12}) \int d^2r_\perp \; e^{-r_\perp^2/ 2\eta_0^2}
t({\mathbf r_\perp}
+{\mathbf y})\tilde p\left( {{2\omega_o}\over { s_0c}}r_\perp \right)\tilde p\left( {{2\omega_e}\over { s_0c}}r_\perp \right) ,
\end{equation} with \begin{equation}{1\over {\eta_0^2}}\equiv {1\over {r_0^2}} -{{2i\omega_p}\over {s_0c}}.
\end{equation}

\subsection{Lateral PSF and numerical results.}

To determine the resolution, we can take the object being viewed to be perfectly transmitting at a single point and perfectly
opaque elsewhere; i.e. $t({\mathbf r_\perp}+{\mathbf y})=\delta^{(2)}({\mathbf r_\perp}+{\mathbf y}).$
We will henceforth also assume $T_{12}<DL$.
Taking the absolute square of $A$, the lateral or transverse PSF is then:
\begin{equation} PSF({\mathbf y})
= \tilde p^2\left( -{{2\Omega_o}\over { s_0c}}y \right)\tilde p^2\left( -{{2\Omega_e}\over { s_0c}}y \right)e^{-y^2/r_0^2}.\end{equation}
The PSF is narrowed relative to that of the standard confocal microscope as
a result of two items: (i) the exponential factor $e^{-y^2/r_0^2}$ and (ii) the factors of 2 inside
each $\tilde p$. The exponential factor is due to the focusing by
the pump lens. The factors of 2, however, appear even if the pump is not focused; they arise in the following way.
In a standard confocal microscope, the Fourier transform of the pupil function arises because $p({\mathbf r_\perp})$ is multiplied by a phase factor $e^{-i{\mathbf k_\perp}\cdot {\mathbf r_\perp}}$ as
the photon propagates in the transverse direction from the point ${\mathbf r_\perp}$ in the focused spot to the axis at the pinhole. Integrating
over ${\mathbf r_\perp}$ then gives the Fourier transform. In our case, both signal and idler exhibit such phase shifts; however,
the pump itself has a radially-dependent phase. The pump photon thus contributes an additional phase factor
equal in size to the sum of the
phases gained by the signal and idler. So the phase is doubled, and the argument of the Fourier transformed pupil function is also doubled. Note that in the degenerate case $\Omega_o=\Omega_e =\omega_p/2$, we can interpret this in the following manner: although we are viewing the signal or idler at frequency $\omega_p/2$, the resolution is being determined by the properties of the pump, which has twice the frequency and thus higher resolution.
Thus, if the pump is ultraviolet, with signal and idler in the visible range, we will end up with visible-light images that have UV-level resolution.

The factors of 2 are the dominant cause of the PSF narrowing unless $r_0$ is comparable to or
smaller than the size of the Airy disk, at which
point the exponential factors begin to introduce additional narrowing.
To get an idea of the sizes of $r_0$ and $R_{airy}$ we can insert some typical values. Assume a
pump of wavelength $\lambda_p=351\; nm$, with signal and idler wavelengths $\lambda_o=\lambda_e=2\lambda_p$, and suppose that all lenses
have radius $a=2\; cm$ and focal length $f=2\; cm$. Then for pump beam of radius $w_0=1\; mm$, we find:
\begin{equation}r_0 = {{\lambda_pf}\over {\sqrt{2}\pi w_0}}\approx 1.6\; \mu m, \quad
R_{airy}= {{1.22\lambda_o f}\over {2a}}\approx 0.43\; \mu m.\label{airy}\end{equation}
Note that it is easy to reduce $r_0$ if necessary: simply place a beam expander into the path of the pump beam.
By increasing the radius of the beam, we fill a larger portion of the focusing lens, thereby effectively increasing its numerical aperture and
allowing the beam to be focused to a smaller spot. We see from eq. (\ref{r0}) that $r_0$ shrinks by the same factor by which the pump radius is expanded. As we increase the pump radius $w_0$, the resolution should remain approximately constant until $r_0$ and $R_{airy}$ are roughly equal; continued increase in $w_0$ beyond this point should then show improving resolution. For
the example above, setting $r_0=R_{airy}$ and using eq. (\ref{airy}) shows that this occurs around $w_0=3.7\; mm$.

\begin{figure}
\centering
\includegraphics[totalheight=2in]{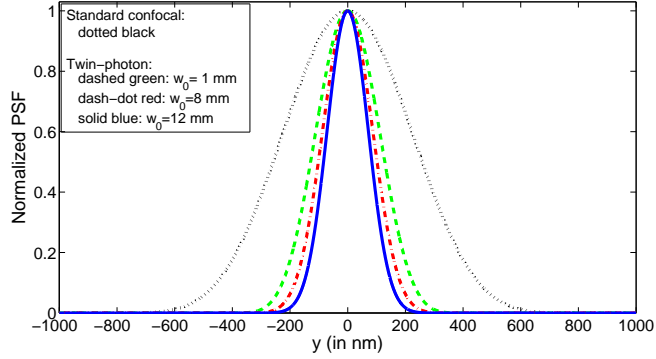}
\caption{\textit{(Color online) Comparison of PSF's for the standard confocal microscope (dotted black) and the twin-photon confocal microscope with
pump radii of $1$ mm (dashed green), $8$ mm (dash-dot red), and $12$ mm (solid blue). The pump is at $351\; nm$. The lenses are in air, with radius $2\; cm$, focal length $2\; cm$, and numerical aperture $NA=1/\sqrt{2}=0.707.$}}\label{compare}
\end{figure}

The calculated PSF of the twin-photon microscope is compared to that of the standard confocal microscope in fig. \ref{compare} for several values of $w_0$, with
all other parameter values as given in the previous paragraph. We see that the PSF decreases in width compared to the standard confocal microscope by $50\%$, $61\%$,
and $68\%$ respectively, for pump radii of $1$ mm, $8$ mm, and $12$ mm.  In the limiting case where the beam
completely fills the focusing lens ($w_0=a=2\; cm$, not shown on the graph), the maximum resolution improvement over the standard confocal microscope is about $77.3\%$. At the opposite extreme of small beam radius, the PSF
remains constant (given by the dashed green curve in fig. 3)
as $w_0$ decreases below about $4$ mm, consistent with the estimate given above. As always, of course, increasing the
numerical aperture of the lenses or increasing the frequency of the light will further improve the resolution.

One further important point should be noted. In confocal microscopy, it is essential that a large angular
range of light be collected by the lenses. So in the setup described here, it is necessary that the downconverted
photons emerge from the crystal with a large angular spread. In noncollinear downconversion, photons of
different frequency are emitted at different angles from the axis, so that the angular spread depends
on the frequency bandwidth. In the case considered in this paper, two factors conspire to increase this spread significantly. First, the spread in frequency is inversely proportional to the thickness of the crystal, which we are assuming to be thin. Second, the tight focusing of the pump beam will cause a further considerable increase in this spread: narrowing the pump in position space will broaden the spread of transverse momentum vectors for both the ingoing and outgoing beams, thus increasing the outgoing angular spread. The focusing of the pump has an effect similar to that of placing a pinhole in the beam;
and as is well known, a small aperture or pinhole will diffract a plane wave to large angles.
In particular, if the lens in the pump has an acceptance half-angle $\alpha$ and numerical aperture $NA=n\sin\alpha$, the pump beam will have an angular spread of $\alpha$ both entering and leaving the focused region. The downconversion will increase the outgoing angular spread by a few additional degrees, depending on the downconversion parameters. So if the objective lens has acceptance angle roughly the same as that of the focusing lens in the pump, then the objective lens will be always be filled. Thus, there will generally be no problem filling a high numerical aperture lens to provide good confocal imaging.

Finally, the apparatus can be simplified as shown in fig. 4. Both outgoing
beams may be sent through the {\it same} lens and separated by a beam splitter after the pinhole.
This not only reduces the number of lenses needed, but should make alignment significantly easier.
In addition, this version has the advantage of increased counting rate, since the full azimuthal angle around the propagation axis
is now covered by the lens.

\begin{figure}
\centering
\includegraphics[totalheight=2in]{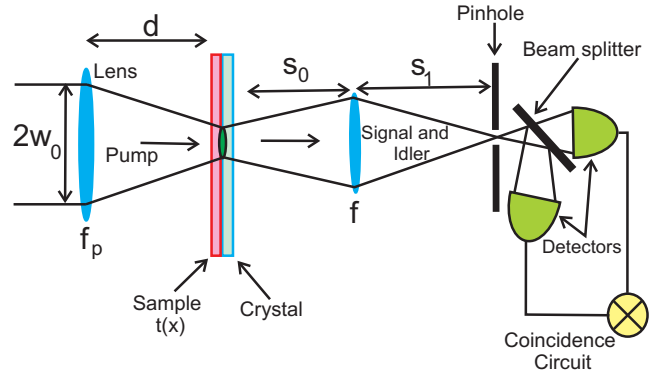}
\caption{\textit{(Color online) Alternate version of twin photon microscope with single lens. The beam splitter after the
pinhole separates the twin photons for coincidence detection.}}\label{setup}
\end{figure}

\section{Discussion}

A large part of the resolution enhancement displayed in the previous section comes from the fact that the illumination is occurring at the smaller pump wavelength; even though we are viewing the image at the longer signal/idler wavelength, the correlated photon production and the coincidence detection together allow us to maintain the inherently higher resolution of the pump. But beyond this, there is a further enhancement that arises if the pump is focused. In
this section, we wish to discuss the latter in more detail. Here, we define the resolution to be the full-width at half-maximum of the image produced when scanning a point object.

Suppose we initially take the illuminating pump beam to be a plane wave. If we hold a point object fixed and scan by moving the detection apparatus (objective lens, pinhole, and detectors) across the sample, the plot of coincidence rate versus scanning position describes a diffractive spot identical to the spot formed by a point object in a standard confocal microscope. This spot defines the area $A_{vis}$ in which a small object is effectively visible to the detectors at each moment, and a plot of its intensity traces out the form of the point spread function for the detection system. This PSF is largest near the particle location and then drops off rapidly with distance. The more rapid the drop-off, the more precisely the position of the object can be localized. The confocal microscope resolves small particles and structures better than a conventional widefield microscope because the drop-off with distance is much sharper.

Now, rather than taking the illumination to be a plane wave, let us focus the pump to a small region.
Imagine at first that we can focus the beam to a mathematical point at the origin. Then during our scan we would only see a non-zero coincidence rate when the particle is precisely at the origin. The region visible to the detection system, $A_{vis}$, is still the same size as before, but there will be nothing to see if the particle is slightly off-axis and therefore not illuminated by the focused pump. Now change the scanning method: hold the detection apparatus fixed and move the particle instead.
As the particle is scanned across the visible region, there is darkness (zero coincidence rate) until the particle crosses the origin, when the coincidence counter lights up briefly, followed by darkness again as the particle moves out of the pump. Therefore, although the optical system through which the outgoing light passes has finite resolution and can localize positions only to within the region $A_{vis}$, the extra information given by the localization of the pump beam to a point allows us to localize the particle position {\it to infinite precision}.

In the real world, diffraction and the laws of quantum mechanics prevent the pump from being focused to a single point, so the best we can do is
focus it to another finite-sized spot $A_{pump}$. But the principle is still the same. The localization of the pump gives extra information about the position of the particle beyond what the outgoing imaging system provides, since now we will see nothing unless the object is within the intersection of $A_{pump}$ and $A_{vis}$. Thus, convolving the illumination amplitude in $A_{pump}$ in with the detection amplitude in $A_{vis}$, we obtain a combined spot which is {\it effectively} smaller. By this we mean that, although the radius of the region visible to the detection system is the same size, the drop off in coincidence rate as the particle moves away from the origin is more rapid due to the combined drop-off of detection and illumination. The location of the half-maximum moves inward toward the center, giving a smaller full-width at half-maximum and thus improved resolution. The standard confocal microscope has improved localization ability over the widefield microscope due to the convolution of one illumination branch and one detection branch; the twin-photon microscope microscope thus goes further, achieving additional localization via the convolution of one illumination branch with the product of {\it two} detection branches.


\section{Conclusions}

We have seen that use of the spatial correlations inherent in the parametric downconversion process allows a type of confocal microscope to be constructed in which the lateral resolution, as measured by the width of the central peak of the point spread function, is greatly improved compared
to the standard confocal microscope with the same optical parameters. The device is essentially a three-photon microscope, requiring the active involvement of the ingoing pump photon, as well as the two twin outgoing photons (the signal and idler) that are actually detected in coincidence.

Because of the requirements imposed by the downconversion process and coincidence counting, the counting rate for a given optical input power will be lower than that
of a traditional confocal microscope. Also, this device lacks the axial (longitudinal) resolution inherent in the
standard confocal microscope. However, the dramatic reduction in the transverse (lateral) size of the
effective confocal region is of clear benefit in many applications, allowing improved ability to optically view
subwavelength nanoscale structures normally visible only in the ultraviolet as the confocal region is scanned over a sample. As a result, the
twin-photon confocal microscope may prove itself a useful complement to the traditional confocal microscope.

\section*{Acknowledgments}
This work was supported by a U. S. Army Research
Office (ARO) Multidisciplinary University Research Initiative
(MURI) Grant; by the Defense Advanced Research Projects Agency (DARPA),
and by the Bernard M. Gordon Center for Subsurface Sensing and Imaging
Systems (CenSSIS), an NSF Engineering Research Center.

\end{document}